\documentclass{aa}

\usepackage{natbib}
\usepackage{graphicx}
\usepackage{amssymb}
\usepackage[dvipsnames]{color}

\begin{document}

\newcommand{\3}{\ss}
\newcommand{\n}{\noindent}
\newcommand{\eps}{\varepsilon}
\newcommand{\be}{\begin{equation}}
\newcommand{\ee}{\end{equation}}
\newcommand{\bl}[1]{\mbox{\boldmath$ #1 $}}
\def\ba{\begin{eqnarray}}
\def\ea{\end{eqnarray}}
\def\de{\partial}
\def\msun{M_\odot}
\def\div{\nabla\cdot}
\def\grad{\nabla}
\def\rot{\nabla\times}
\def\ltsima{$\; \buildrel < \over \sim \;$}
\def\simlt{\lower.5ex\hbox{\ltsima}}
\def\gtsima{$\; \buildrel > \over \sim \;$}
\def\simgt{\lower.5ex\hbox{\gtsima}}

\title{First supernovae in dwarf protogalaxies}

\author{E.~O.~Vasiliev,
\inst{1,3}
E.~I.~Vorobyov,
\inst{2,3}
          \and
Yu.~A.~Shchekinov
\inst{4,5}
}

   \offprints{E.~O.~Vasiliev} 
   
   \institute{Tartu Observatory, 61602 T\~oravere, Estonia 
\email{eugstar@mail.ru}
\and
Institute for Computational Astrophysics, Saint Mary's University,
Halifax, B3H 3C3, Canada
\email{vorobyov@ap.smu.ca}
\and
Institute of Physics, Southern Federal University,
Stachki Ave. 194, Rostov-on-Don, 344090 Russia
\and
Department of Physics, Southern Federal University,
Sorge St. 5, Rostov-on-Don, 344090 Russia
\email{yus@phys.rsu.ru}
\and 
Special Astrophysical Observatory, 
Nizhny Arkhyz, 
Karachai-Cherkessia, 369167 Russia   
}

   \date{}

% \authorrunning{Vasiliev, Vorobyov, Shchekinov}
% \titlerunning{First supernovae in dwarf protogalaxies}

\abstract
{This paper is motivated by the recent detection of many extremely metal-deficient 
($[\rm{Fe/H}]<-3$) stars in the Milky Way.}
{We explore numerically the chemical, thermal, and dynamical evolution of a shell formed 
by a high-energy supernova explosion ($10^{53}$~erg) in dwarf protogalaxies with 
total (dark matter plus baryonic) mass $10^7~M_\odot$ at a redshift $z=12$. 
We consider two initial configurations for the baryonic matter, one without rotation and 
the other having the ratio of rotational to gravitational energy $\beta=0.17$. The (non-rotating) 
dark matter halo is described by a quasi-isothermal sphere. The latter choice is motivated 
by recently proposed mechanisms for rapid flattening of a central cuspy 
region in dwarf protogalaxies. }
{We use a finite-difference numerical hydrodynamics code to 
simulate supernova explosions in dwarf protogalaxies with axial symmetry.
The advection is treated using a third-order piecewise parabolic scheme.
The heating and cooling processes in the gas are taken into account by solving numerically 
the rate equations for main atomic, molecular and ionic species in the primordial gas.}
{We find that the dynamics of the shell is different in protogalaxies with and without rotation. 
For instance, the Rayleigh-Taylor instability in the shell develops faster in 
protogalaxies without rotation. The fraction of a blown-away baryonic mass
is approximately twice as large in models with rotation ($\sim 20\%$) 
than in models without rotation. We argue that these differences are 
caused by different $initial$ gas density profiles in non-rotating and rotating protogalaxies. 
On the other hand, the chemical evolution of gas in protogalaxies with 
and without rotation is found to be similar. The relative number 
densities of molecular hydrogen and HD molecules in the cold gas 
($T \le 10^3$~K) saturate at typical values of $10^{-3}$ and 
$10^{-7}$, respectively. The saturation times in models with 
rotation are somewhat longer than in models without rotation.
The clumps formed in the fragmented shell move with velocities that 
are at least twice as large as the escape velocity. The mass of the
clumps is $\sim 0.1-10~\msun$, which is lower 
than the Jeans mass. We conclude that the clumps are pressure 
supported.}
{A supernova explosion with energy $10^{53}$~ergs destructs 
our model protogalaxy. The clumps formed in the fragmented shell 
are pressure supported. We conclude that protogalaxies with total mass $\sim 10^{7}~M_\odot$ 
are unlikely to form stars due to high-energy supernova explosions of the first stars.}
\keywords{cosmology: early universe -- galaxies: formation --
ISM: molecules -- stars: formation -- shock waves}

\maketitle

\section{Introduction}

The detection of extremely metal-poor 
stars in our Galaxy with the iron abundance equal to or less than $10^{-3}$ 
of the solar value \citep{beers,christlieb02}
motivated the scientists to put forward possible scenarios 
for the formation of such stars. According to \citet{tsuji99}
extremely metal-poor (EMP) stars form in a dense shell produced by
Type II supernova explosions of the first stars and accrete metals from 
the surrounding 
medium during the subsequent evolution. The formation of EMP stars
is made possible by fragmentation of the primordial gas in a supernova 
shell due to efficient cooling by molecular hydrogen and HD molecules. 
Indeed, the formation of molecular hydrogen increases significantly  
in the wake of a strong shock wave \citep{suchkov83,sk87}.
Moreover, at temperatures lower than 
$150~$K, HD molecules become a much more efficient coolant 
than molecular hydrogen \citep[e.g.,][]{sh86}.

Subsequent studies by many authors 
\citep[see review by][]{nishi99,salva04,machida,greif} 
have shown that the likelihood for the formation of EMP stars 
in supernova-driven shells is sensitive to both 
the mass of the dark matter halo ($M_{\rm h}$) in a primordial galaxy
and the assumed type of supernova explosions.
For instance, \citet{nishi99} have shown that the shell fragmentation
due to Type~II supernova explosions ($10^{51}$~erg)
is possible only in dark matter halos with total mass 
$M_{\rm h} \ga {\rm a~few} \times 10^7~\msun$.
Galaxies with $M_{\rm h} \la 10^7~\msun$ may lose its baryonic matter
before it has time to fragment and form stars \citep{ferrara98}. 
The likelihood for this blow-away scenario increases if the first stars explode
as pair-instability supernovae with the energy release of the order of
$10^{53}$~erg \citep{bromm03,greif}. However, these numerical results 
were confronted by \citet{machida}, who presented a semi-analytic model for 
the evolution of a gas shell produced by supernova explosions with energy 
$10^{51}-10^{52}$~erg. They took into account the H$_2$ and HD 
chemistry and found that supernova explosions can induce fragmentation
of the gas shell and formation of EMP stars in dark matter halos 
with $M_{\rm h} = {\rm a~few} \times 10^6~\msun$. 
The typical mass of the fragments in their model is about $1.0~\msun$, 
which is consistent with the result by \citet{uehara00}
and similar to the masses of the observed low-metallicity stars
\citep{christlieb02}. 

The likelihood for the formation of EMP stars depends also on
the radial distribution of gas in a primordial galaxy {\it prior} to supernova 
explosions. This distribution can (partly) be determined by HII regions around
the first stars. 
For instance, \citet{whalen04} considered the formation of the HII 
region in a low-mass dark matter halo with mass $10^6~\msun$ and
found that the gas is efficiently blown away by a supersonic shock wave
accociated with the R-type ionization front.
Similar results were obtained by \citet{kitayama04} for the same 
mass of the dark matter halo. They have also
considered more massive dark matter halos and concluded that the 
HII region is confined within the virial radius in halos with masses 
$\simgt 10^7~\msun$ (the ionization front is of the D-type). 
More recently, \citet{kitayama05} studied numerically the 
destruction of dark matter halos due to the formation of HII 
regions due to radiation of the first stars and subsequent supernova explosions.
Using a one-dimensional Lagrangean hydrodynamics code, they found that
the SN shock wave remains well inside the virial radius for dark matter 
halos with masses larger than $10^7~\msun$.
On the other hand, in low-mass dark matter halos with
masses of the order of $10^6~\msun$ the shock wave propagates outside 
the virial radius. These results  indicate that the efficiency of 
halo destruction (and, by implication, the efficiency for the formation 
of EMP stars) are determined not only by the explosion energy but also 
by the gas density distribution
and radiative feedback from the first stars prior to their explosion.

It now becomes evident that the formation of EMP stars due to fragmentation
of supernova driven shells is a complicated phenomenon that depends on 
a variety of physical conditions in a host protogalaxy, 
which may vary from allowing
star formation to shutting it off completely. In such circumstances, the
construction of increasingly more sophisticated numerical models is justified.
The majority of abovementioned studies are based on semianalytic models
\citep[e.g.,][]{salva04,machida} or on one-dimensional numerical 
hydrodynamic simulations 
\citep[e.g.,][]{kitayama05}. In the framework of multi-dimensional 
numerical simulations, the priority was often given to a careful 
investigation of dynamical processes leading to the destruction of a 
dwarf protogalaxy by supernova explosions, neglecting the chemical processes
that control the gas thermal properties. In the present paper, 
we perform axially symmetric numerical hydrodynamics simulations 
of high-energy supernova explosions ($10^{53}$~erg) in a model 
dwarf protogalaxy with total (dark matter plus baryonic) mass 
$1.7\times 10^7~M_\odot$. We give a careful treatment to the heating and 
cooling processes and chemical evolution of main molecular, ionic, and atomic 
species. We seek to determine the effect of galactic rotation on the dynamical and
chemical evolution of a supernova-driven shell.
The dark matter halo profile in our models is defined by a modified isothermal sphere 
\citep[as observed in local dwarf galaxies,][]{burkert},
%Burkert 1995), 
rather than by a cuspy profile \citep{nfw}.
%(Navarro et al. 1997).
%\textcolor{blue}{Very recently \citet{greif} have performed a
%study of first SN
%explosions in three-dimensional cosmological simulation. They considered %SN explosion 
%with energy $10^{52}$~erg in small dark halo witn 
%$M=5\times 10^5~\msun$ 
%after almost fully its destruction by ionizing radiation,
%so the gas profile before the SN explosion is flat with number density 
%$n\simeq 0.5$~cm$^{-3}$. }

The paper is organized as follows. 
In Sections 2 and 3 we describe the model and numerical techniques, 
respectively; in Section 4 we discuss several neglected processes;
in Section 5 we present results of our numerical simulations.
The discussion and conclusions are given in Sections 6 and 7, respectively.
Throughout the paper we assume a $\Lambda$CDM cosmology with the parameters
$(\Omega_0, \Omega_{\Lambda}, \Omega_m, \Omega_b, h ) =
(1.0,\ 0.76,\ 0.24,\ 0.041,\ 0.73 )$ 
as inferred from the Wilkinson Microwave Anisotropy Probe (WMAP),
and deuterium abundance $2.78\times 10^{-5}$,
consistent with the most recent measurements \citep{spergel06}.

\section{Model protogalaxy}

Our model protogalaxy consists of a baryonic component surrounded 
by a spherical dark matter halo. We assume that the dark 
matter halo profile is spherically 
symmetric and is determined by a modified isothermal sphere
\be
 \rho_h(r) = {\rho_0 \over 1 + (r/r_0)^2},
 \label{halo}
\ee
where $r_0$ and $\rho_0$ are the core radius and central density, respectively.
Cosmological N-body simulations suggest that dark matter halos (upon their formation)
have radial profiles that are cuspy in the central regions and scale as $r^{-3}$ 
at large radii \citep{nfw}. On the other hand, modeling of the rotation curves in nearby dwarf 
galaxies indicates that the dark matter profiles have a flat 
central region and a tail that scales as $r^{-2}$ \citep{burkert}.
It is poorly known how and when the transition from cuspy to flat 
dark matter halos occurs. Several possible mechanisms that can 
facilitate transformation of the profile include random velocities of baryons due to
a star formation feedback \citep{mashch},
initial background perturbations of the dark matter \citep{dor06}
and angular momentum transfer from baryons to the dark matter \citep{tonini}.
In particular, \citet{mashch} suggest that random bulk motions of 
gas in small primordial galaxies can flatten the central dark matter 
cusp on relatively short timescales, $\sim 10^8$~yrs\footnote{For the 
adopted cosmological model, this timescale corresponds to 1/4
of the comoving age of the universe at $z=12$.}. Motivated by this 
finding, we assume that dark matter halos in protogalaxies at $z=12$ 
have already flattened and attained a configuration
described by equation~(\ref{halo}). 

We calculate $r_0$ and $\rho_0$ as a function of the halo
mass based on the empirical profile for nearby dwarf 
galaxies \citep{burkert,Silich} 
extrapolated to the early universe by \citet{fujita}
\be
 r_0 = 8.9\times10^{-6} \ {\rm kpc} \ M_h^{1/2} h^{1/2} \Omega_m^{-1/3}(1+z)^{-1},
\ee
\be
 \rho_0 = 6.3\times 10^{10} \ {\rm \msun kpc^{-3}} \ M_h^{-1/3} h^{-1/3} \Omega_m (1+z)^3.
\ee
The dark halo mass ($M_{\rm h}$) in our numerical simulations is set to $10^7~M_\odot$, which,
at a redshift of $z=12$, corresponds to $3\sigma$ perturbations in 
the $\Lambda$CDM model for the parameters determined from the third 
year WMAP data. According to \citet{t97} and \citet{sv06}, 
protogalaxies with dark matter halos of order $10^7~M_\odot$ are
expected to cool down efficiently. 
The virial radius of our model protogalaxy is $r_{\rm v} = 520$~pc 
\citep[for relations between the virial parameters see e.g.,][]{cfrev}.

The initial distribution of the total gas density ($\rho_{\rm g}$) 
is found by solving numerically the steady-state momentum equations in 
cylindrical coordinates $(z,r)$
\begin{eqnarray}
{u^2_\phi \over r} &=& {1 \over \rho_{\rm g}} {\partial p \over 
\partial r} + {\partial \Phi_{\rm h} 
\over \partial r}, \label{equilib1} \\
{1 \over \rho} {\partial p \over \partial z} &=& - {\partial 
\Phi_{\rm h} \over \partial z},
\label{equilib2}
\end{eqnarray}
where $p=\rho_{\rm g} c^2_{\rm s}$ is the gas pressure, $c^2_{\rm s}=R T/ \mu$ is the sound speed,
$\mu$ is the mean molecular weight, $\Phi_{h}$ is 
the gravitational potential of the dark matter halo, and $u_{\phi}$ is the rotation 
(azimuthal) velocity of gas. The initial value of $\mu$ is set to 1.22, 
which corresponds to a gas of neutral hydrogen and helium 
with abundances by mass $X=0.76$ and $Y=0.24$, respectively.
The gas is assumed to be {\it initially} isothermal at a virial 
temperature $T_{\rm vir}=5900$~K. 

In order to solve equations~(\ref{equilib1}) and (\ref{equilib2}) for the initial 
gas density distribution, we have to know the initial rotation 
velocity of gas, $u_{\rm \phi0}$. 
We determine $u_{\rm \phi0}$ in two steps. First, we calculate 
the circular velocity of gas $u_{\rm circ}$, which is determined 
exclusively by the gravitational potential of the dark matter halo
(neglecting the contribution from gas pressure gradients)
\be
u^2_{\rm circ}= r {\partial \Phi_{\rm h} \over \partial r}.
\ee
Then we introduce a parameter $\alpha<1$, which 
measures the relative input of rotation to 
the total support against gravity. The rotation velocity is finally determined as 
$u_{\rm \phi 0}=\alpha \, u_{\rm circ}$. In the following text, 
we consider two models: model~1 without rotation and model~2 with rotation, 
for which we choose $\alpha=0.4$. The resulted total rotational energy inside 
the virial radius is $E_{\rm rot}=\int u_{\phi0}^2 \, \rho \, dV \approx 0.31\times 10^{51}$~erg.
The corresponding ratio $\beta$ of rotational to gravitational energy is equal to 0.17.
We note that $\beta \simeq \alpha^2$ because 
$E_{\rm rot}/E_{\rm gr} \simeq u_{\rm \phi}^2/u_{\rm circ}^2$, where $E_{\rm gr}$
is the gravitational energy.
%which is about $17\%$ of the corresponding total gravitational energy,
%$E_{\rm gr}\approx 1.97\times 10^{51}$~erg. 
 We note that our model galaxy
has a modest amount of rotation and most support against gravity 
comes from gas pressure gradients.

The numerical procedure for solving the steady-state equations~(\ref{equilib1})
and (\ref{equilib2}) to obtain the equilibrium gas density distribution is given in \citet{vorob04}.
There is, however, one important difference. The total masses of gas
and dark matter halo in protogalaxies are linked through the relation 
$M_{\rm g}= M_{\rm h} (\Omega_b / \Omega_m )$.
In order to satisfy this relation, we iterate solutions until the resulted gas density 
distribution has attained the mass of about $1.7\times 10^6~M_\odot$
inside the virial radius.

\begin{figure}
  \resizebox{\hsize}{!}{\includegraphics{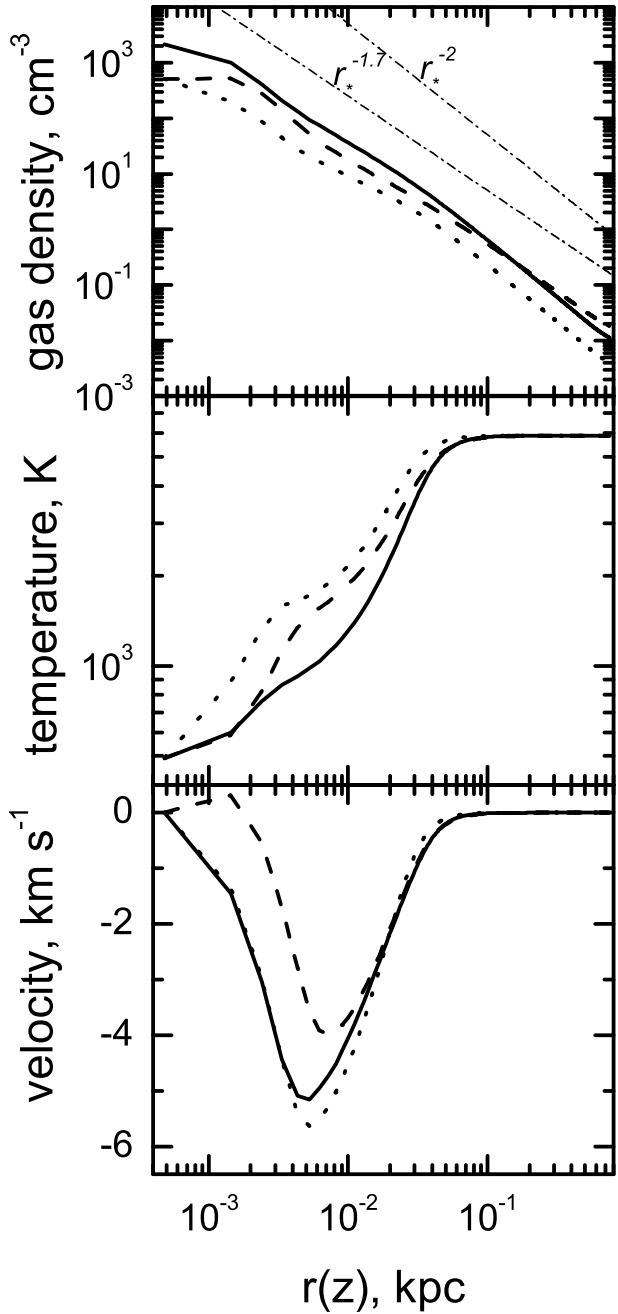}}
      \caption{
      Profiles for the gas density (top panel), temperature (middle panel), and 
      infall velocity (bottom panel) in the non-rotation
      model~1 (solid line)  and rotating model~2 (dashed and dotted lines).
      In particular, the dashed and dotted lines show the radial 
      and vertical profiles in model~2, respectively.}
  \label{Fig1}
\end{figure}

Once the equilibrium gas density distribution is constructed, we let our model galaxies evolve 
out of equilibrium. The inner densest regions are characterized by largest cooling rates, and
they cool and contract on much shorter time scales than the outer regions.  
We stop this process when the temperature in the center drops below 500~K, correspondingly the
density increases about 50 times in comparison with the central equilibrium value.
This happens at $t\ga 1.0$~Myr in model~1 and $t\ga 1.4$~Myr in model~2.
The resulted distributions 
of gas density (top panel), temperature (middle panel), and infall velocity (bottom panel)
in model~1 (solid line) and model~2 (dashed and dotted lines) are shown in Figure~\ref{Fig1}. 
More specifically, the dashed and dotted lines show the radial and vertical
profiles in model~2, respectively. We note that in the non-rotating 
model~1 the distributions are identical in all directions.  

The infall velocities in Figure~\ref{Fig1} indicate that only the inner regions of our 
model galaxy $r\la 50$~pc are driven out of equilibrium, while the rest of the galaxy is intact.
This out-of-equilibrium region remains confined to the inner 100~pc, even if we extend our numerical
integration for another several million years. As a result, the radial gas density profiles 
in model~1 and 2 are distinct throughout the bulk of the galaxy. In particular,
the {\it radial} gas density profile in model~2 (with rotation) is shallower ($\propto r^{-1.7}$) 
than in model~1 without rotation ($\propto r^{-2.0}$), except for the innermost region. 
The difference in the central gas density
between the models with and without rotation can reach a factor of three.
However, the masses enclosed inside the virial radius ($\approx 520$~pc) are 
approximately the same in both models. The vertical gas distribution in model~2 (dotted line)
is characterized by progressively smaller values (at same radii) than 
in model~1 (solid line). This implies that supernovae remnants in 
rotating protogalaxies are expected to expand to a larger distance 
in the vertical direction, which may result in a partial loss
of gas by a parent protogalaxy. 

Once the central gas temperature drops below 500~K, 
we release $10^{53}$~erg of thermal 
energy in the central sphere with radius 5~pc. Such energetic supernovae 
are expected to result from the pair-instability explosion 
of massive metal-free stars \citep{heger}.
This approach is supposed to reasonably
mimic the formation of a first star and its subsequent explosion.
The gas temperature in the central sphere is set to $2.5\times 10^4$~K, typical for HII regions,
and gas density is adjusted to maintain the pressure balance between the interior and immediate
exterior of the sphere. We assume
that this sphere was formed by a collective action of stellar 
wind and ultraviolet radiation from a massive star (the supernova 
progenitor) that had resided in the galactic center 
before the explosion (we skip this phase in our numerical simulations). 
Since the physical size of a shock front at initial stages is considerably smaller than
our numerical resolution ($\sim 1$~pc), we allow for the gas to initially evolve adiabatically
and switch on cooling only after $6\times 10^4$~yr of evolution. 
In what follow we will loosely define this hot sphere as 
a supernova ejecta, although it differs physically from the ejecta 
in realistic supernovae explosions. The initial composition of species inside
the sphere corresponds to a fully ionized gas.

\section{Numerical model}

\subsection{Main equations}
\label{maineq}
The dynamics of the gaseous component in our model protogalaxy 
is followed by numerically solving a usual set of hydrodynamic equations
in cylindrical coordinates ($z,r,\phi$)
\begin{eqnarray}
{\partial \rho_{\rm g} \over \partial t} &+& {\bl \nabla} \cdot (\rho_{\rm g} 
{\bl u} ) = 0, \label{hydro1} \\ 
{\partial \over \partial t } (\rho_{\rm g} {\bl u}) &+& {\bl \nabla} \cdot (\rho_{\rm g}
{\bl u} \cdot {\bl u}) = - {\bl \nabla} p - \rho_{\rm g}{\bl \nabla} \Phi_{\rm h}, \label{hydro2} \\ 
{\partial \epsilon \over \partial t} &+& {\bl \nabla} \cdot (\epsilon \, {\bl u}) =
-p ({\bl \nabla} \cdot {\bl u}) - \Lambda,
\label{hydro3}
\end{eqnarray}
where ${\bl u}={\hat {\bl z}} u_{z} + {\hat {\bl r}} u_{r} +{\hat \phi} u_\phi$ is
the gas velocity in cylindrical coordinates, $\bl \nabla={\hat {\bl z}} \partial / \partial z + 
{\hat {\bl r}}
\partial / \partial r$ is the spatial derivative in cylindrical coordinates
modified for the adopted axial symmetry, $\Lambda$ is
the cooling rate, and $\epsilon$ is the internal energy density.
The hydrodynamic equations are closed by an ideal equation of state $p=(\gamma-1)\epsilon$,
where $\gamma$ is set to 5/3. A (small) amount of numerical viscosity is added to the code
(according to the usual prescription of von Neumann and Richtmyer) in order to smooth out shock 
and discontinuities over two computational zones.

Equations~(\ref{hydro1}-\ref{hydro3}) are solved using a finite-difference,
operator-split code, which applies the consistent transport technique introduced
by \citet{Norman}. The code performs well on the standard test suite 
including the Sedov point explosion test. The advection is treated 
using a third-order piecewise parabolic advection scheme \citep{ppm}.
This scheme is known to be more accurate in handling
the shocks and contact discontinuities than a commonly
used van Leer advecion scheme \citep[see e.g.,][]{zeus}.
%A predictor-corrector method is used to update the internal energy 
%due to compressional heating and cooling terms. This method, though
%less efficient than Newton - Raphson iterations, is found to be more 
%stable for strongly nonlinear problems. 
The computational domain has 
a size of 750~pc in both the vertical ($z$) and horizontal
($r$) directions. Cosmological effects (expansion and bias) are not expected
to significantly influence the results on such spatial scales.
We assume an equatorial symmetry for simplicity. The numerical resolution
is $780\times 780$ grid zones, which corresponds to a spatial
resolution of $\approx 1.0$~pc in both coordinate directions. 
In our simulations self-gravity of the gas is neglected and the dark matter 
potential is fixed. 
In a few test runs, we included the self-gravity of the gas but found it
dynamically unimportant.

\subsection{Cooling processes and chemical reactions}

The gas component of our model protogalaxy
consists of a standard set of species: H, He, H$^+$, H$^-$, H$_2$, 
H$_2^+$, D, D$^+$, and HD. We also make use of the charge conservation 
law for electrons. The initial number densities of neutral hydrogen and 
helium  relative to the total number density of gas $n_{\rm g}=\rho_g/(\mu m_{\rm H})$ 
are $x[{\rm H}]=0.9999 \times 0.76$ 
and $x[{\rm He}]=0.24$, respectively. The initial number densities of ionized hydrogen, molecular
hydrogen, and HD molecules relative to $n_{\rm g}$ are $x[{\rm H^+}]=10^{-4}$, 
$x[{\rm H_2}]=10^{-6}$ and $x[{\rm HD}]=10^{-10}$, respectively. 
The initial relative number density of deuterium is equal to 
the cosmological value. The relative number densities of 
the other species in our sample are set initially to a negligible value.

The cooling rates are computed separately for temperatures below and above 
$2\times 10^4$~K. In the low-temperature regime, the cooling rate 
$\Lambda$ in the energy equation~(\ref{hydro3}) includes typical 
radiative losses in the primordial plasma. 
Cooling due to recombination and collisional excitation of atomic 
hydrogen is taken from \citet{cen92}. 
It is worth mentioning \citet{galli98} H$_2$ cooling function almost concides
with the total cooling function including H--H$_2$ and e$^-$--H$_2$ collisions
for $x_e < 10^{-4}$ \citep{glover08}, which is the case in our simulation. 
Cooling due to molecular 
hydrogen is taken from \citet{galli98} and modified for 
the temperature regime near the CMB radiation temperature 
\citep{varsh76,puy93,flowerh2}. Cooling due to HD molecules is 
computed according the prescription given in 
\citet{flower00} and \citet{lipovka05}. In the high-temperature regime, 
the cooling rates for zero metallicity are taken from \citet{sd93}.

%%%%%%%%%%%%%%%%%%%%%%%%%%%%%%%%%%%%%%%%%%%%%%%%%%%%%%%%%%%%%%
\begin{table}[!ht]
\caption{Chemical reaction rates}
\begin{tabular}{lc}
\hline
reaction              & Reference \\
\hline
${\rm H^+   +   e^-  \to H     +   h\nu}$    &\citet{galli98}    \\  %H1
${\rm H     +   e^-  \to H^-   +   h\nu}$    &--                 \\  %H3
${\rm H^-   +   h\nu \to H     +   e^-}$     &--                 \\  %H4
${\rm H^-   +   H    \to H_2   +   e^- }$    &--                 \\  %H5
${\rm H^-   +   H^+  \to H_2^+ +   e^- }$    &--                 \\  %H6
${\rm H^-   +   H^+  \to 2H            }$    &--                 \\  %H7
${\rm H^+   +   H    \to H_2^+ +   h\nu}$    &--                 \\  %H8
${\rm H_2^+ +   H    \to H_2   +   H^+ }$    &--                 \\  %H10
${\rm H_2^+ +   e^-  \to 2H            }$    &--                 \\  %H11
${\rm H_2   +   H^+  \to H_2^+ +   H   }$    &--                 \\  %H15
${\rm H_2   +   e^-  \to 2H    +   e^- }$    &--                 \\  %H17
${\rm H_2   +   H    \to 3H            }$    &\citet{maclow86}   \\  %kdiss
${\rm H     +   e^-  \to H^+   +   2e^-}$    &\citet{abel97}     \\  %A1
${\rm H^-   +   e^-  \to H     +   2e^-}$    &--                 \\  %A14
${\rm H^-   +   H    \to 2H    +   e^- }$    &\citet{sk87}       \\  %S17
${\rm D^+   +   H_2  \to HD    +   H^+ }$    &\citet{galli98}    \\  %D8
${\rm H^+   +   HD   \to H_2   +   D^+ }$    &--                 \\  %D10
\hline
\end{tabular}%
\label{table1}
\end{table}
%%%%%%%%%%%%%%%%%%%%%%%%%%%%%%%%%%%%%%%%%%%%%%%%%%%%%%%%%%%%%%

We assume that  our atomic, ionic, and molecular species are collisionally
coupled to each other and share a common velocity field, which eliminates the need for 
solving separate equations of motion for each species. 
Hence, in addition to the usual hydrodynamic
equations~(\ref{hydro1}-\ref{hydro3}), we have to solve the continuity and rate
equations for the mass densities ($\rho_{\rm i}$) of each of the species 
\be
{\partial \rho_{\rm i} \over \partial t} + {\bl \nabla} \cdot (\rho_{\rm i} {\bl u}) =
k_{\rm jks} \rho_{\rm j} \rho_{\rm k} - k_{\rm ki} \rho_{\rm k} \rho_{\rm i},
\label{chem}
\ee
where the right-hand terms are the sources and sinks due to chemical 
reactions. The list of reactions and references to the corresponding 
formation ($k_{\rm jl}$) and destruction ($k_{\rm ki}$) rates   
are given in Table~\ref{table1}. We note that we do not solve equations~(\ref{chem}) 
for D$^{+}$ and D, since the ratio of ionized versus neutral deuterium densities can be 
implicitly derived from the ratio of ionized versus atomic hydrogen 
densities \citep{vs03}. We have also neglected the helium kinetics.

The hydrodynamic equations~(\ref{hydro1}-\ref{hydro3}) and the chemical reaction 
network~(\ref{chem}) are numerically coupled using the following strategy. First, the 
cooling rate $\Lambda$ is computed and the global hydrodynamic time step $\Delta t_{\rm h}$
is derived as
\begin{equation}
{1\over \Delta t_{\rm h}} = {1\over \Delta t_{\rm CFL}} + {1\over \Delta t_{\rm cool} },
\end{equation}
where $\Delta t_{\rm CFL}$ is a usual hydrodynamic time step due to the Courant-Friedrichs-Lewy 
condition and $\Delta t_{\rm cool} = 0.1 \epsilon/\Lambda$ is a characteristic cooling time.
Note that $\Delta t_{\rm h}$ should be minimized over all computational cells.

Once the global hydrodynamic time step is calculated, the solution of 
equations (\ref{hydro1})-(\ref{chem}) proceeds as follows.
First, we solve the continuity and momentum equations
for the total gas density $\rho_{\rm g}$ (eqs.~(\ref{hydro1}) and (\ref{hydro2}), respectively)
using a technique described in Section~\ref{maineq}.
The solution of the chemical reaction network~(\ref{chem}) and energy balance equation~(\ref{hydro3})
 is split into two parts.
The update of internal energy density ($\epsilon$) and mass densities of each species 
($\rho_i$) due to advection is done by solving the following continuity equations 
\begin{eqnarray}
{\partial \rho_{\rm i} \over \partial t}  &+& {\bl \nabla} \cdot (\rho_{\rm i} {\bl u}) = 0, \\
{\partial \epsilon \over \partial t} &+& {\bl \nabla} \cdot (\epsilon \, {\bl u}) = 0.
\end{eqnarray}
The solution technique is exactly the same as for equation~(\ref{hydro1}).

The remaining update of $\rho_i$ due to chemical reactions 
is combined with the update of $\epsilon$ due to cooling 
and compressional heating
\begin{eqnarray}
{\partial \rho_{\rm i} \over \partial t} &=&
k_{\rm jk} \rho_{\rm j} \rho_{\rm k} - k_{\rm ki} \rho_{\rm k} \rho_{\rm i}, \label{rate} \\
{\partial \epsilon \over \partial t} & =&
-p ({\bl \nabla} \cdot {\bl u}) - \Lambda.
\label{coolheat}
\end{eqnarray}
To improve the accuracy, we use a predictor-corrector scheme. At each step of 
the predictor-corrector method, equations~(\ref{rate})
are integrated using a fifth-order Runge-Kutta-Carsp method with the adaptive step 
size control \citep{numrec}. The global integration time step is determined by 
$\Delta t_{\rm h}$.  
We note that in the high-temperature regime ($T > 2\times 10^4$~K) the mass densities 
of atomic and ionized hydrogen are computed adopting a local thermal equilibrium. 
More specifically, the ionization and recombination rates for hydrogen are set equal.
The mass densities of the other four species (H$^-$, H$_2$, H$_2^+$, 
and HD) are set to a negligible value. 

%\subsection{Supernova explosion}
%
%
%\textcolor{red}{\bf Describe changes in the initial central sphere}

\section{Neglected processes}

In our numerical simulations, we have neglected several physical processes
that may be important in some dwarf protogalaxies. For instance, a massive
star -- a likely supernova progenitor -- emits enormous
ultraviolet (UV) flux that ionizes the surrounding gas and forms the HII
region. The radiative pressure of UV photons from massive stars is so strong
that it can blow the gas away from the center of a low-mass ($\sim 10^6~M_\odot$
at $z= 20$) protogalaxy \citep{whalen04}. Our model protogalaxy has a more massive
dark matter halo, $M = 10^7~\msun$, but is located at a lower redshift, 
$z=12$. An HII region formed inside such a protogalaxy is expected 
to be confined within $10$~pc. This estimate is obtained 
by assuming the emission rate of ionizing 
photons from the first stars to be $10^{50}$~s$^{-1}$ \citep{schaerer02}
and background density $n=10$~cm$^{-3}$. It is worth noting however that the size of HII region depends on 
the dark matter profile of a protogalaxy: it is expected to be larger in 
protogalaxies with a flat density profiles, and lower in the cuspy 
profiles. Nevertheless, in order to avoid an overestimate of the effects of the HII region,
we set its size in our numerical simulations to 5.0~pc.

%% The resulted size of the HII region is comparable with the size of 
%% the central near-constant-density core in our model. 
Shock waves, another source of ionizing radiation \citep{shull79},
are efficient only in the adiabatic phase of expansion, 
when the temperature is much higher than $10^4$~K. They cannot 
influence significantly the formation of molecules at later times.

Another feedback process that can affect the chemistry of gas
is the radiation in the ($11.18-13.6$)~eV band, which photo-dissociates 
H$_2$ molecules. Although a photo-dissociative region created by the radiation
of a massive star can be quite large \citep{haiman97},
the typical relative molecular hydrogen density  in this region 
is quite small, about $10^{-6}$ (taken as an initial value in 
our numerical simulations).  Obviously, the primary source of 
dissociative photons ceases to exist after the supernova explosion and 
we expect that the production of dissociative photons by the hot gas 
of a bubble created by the explosion is negligible. 
Moreover, H$_2$ molecules cannot be photo-dissociated inside cold and dense
regions of a supernova-driven shell due to the self-shielding effect \citep{draine96}.
%(Draine \& Bertoldi 1996). 
All this argues against photo-dissociation  as an important mechanism 
for H$_2$ destruction {\it after} the supernova explosion.

A positive feedback on the abundance of H$_2$ molecules 
can come from the X-ray photons produced by supernova explosions
\citep{ferrara98,haiman97}.
However, other coacting mechanisms are likely to be more important.
For instance, it is well established that the 
H$_2$ formation is very efficient behind strong shock waves \citep{suchkov83,sk87}.
As a result, the relative density of H$_2$ 
molecules increases very rapidly below $10^4$~K and saturates 
at a so-called  ''universal'' value of about  
$x[{\rm H_2}] \sim (1-2)\times 10^{-3}$ at  temperature 
$\sim 3\times 10^3$~K \citep{ohhaiman}. Thus, the X-ray photons have 
only a minor influence on the H$_2$ abundance in a supernova-driven 
shell but they can strongly influence the H$_2$ abundance
in the gas ahead the shell. However, this is not expected to 
change the picture as a whole because shock waves created by superovae with
the energy release of order $10^{53}$~erg are strong.

To summarize, we do not expect that these processes are significant for
our model protogalaxy. In a future work, we plan to include 
radiative transfer in 
the numerical hydrodynamics simulations to check the validity of our assumptions.

\section{Results}

\begin{figure*}
  \resizebox{\hsize}{!}{\includegraphics{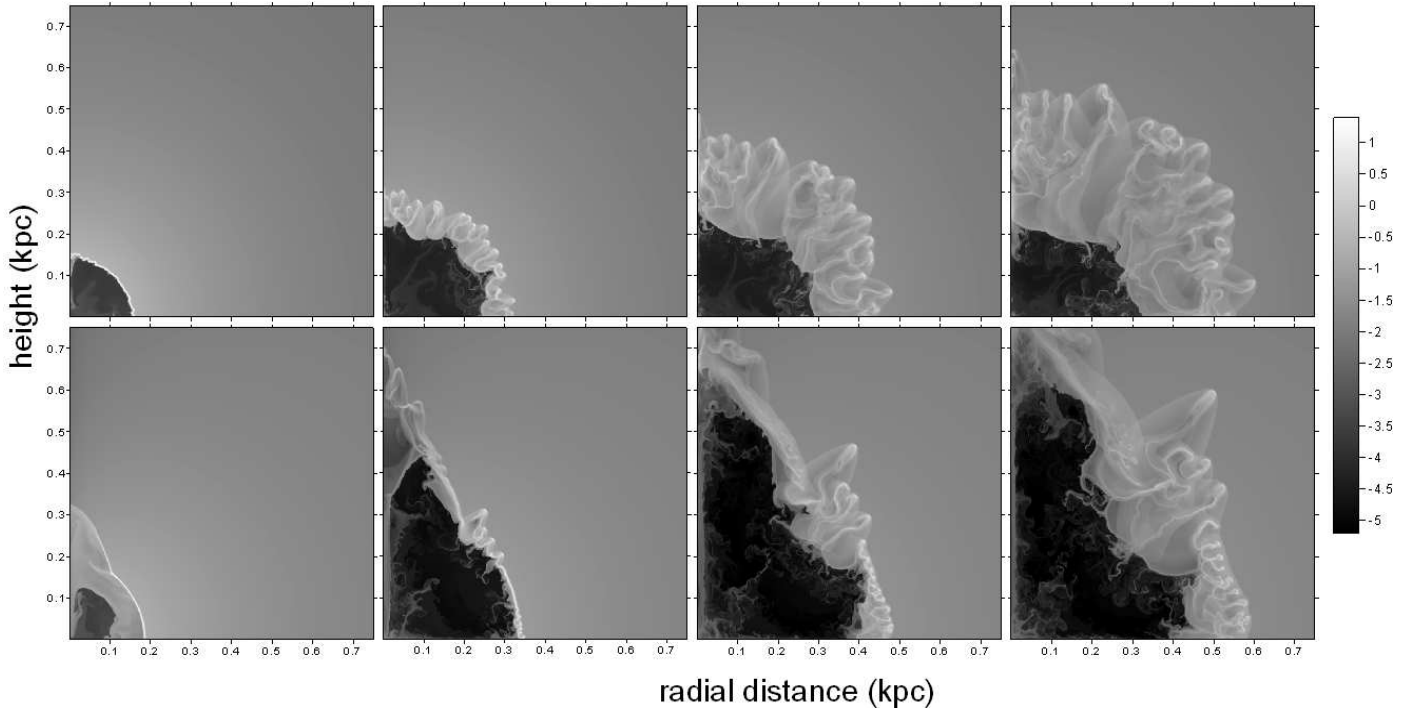}}
      \caption{
               Logarithmic distribution of 
               density (in cm$^{-3}$) at 
               $t = 1,~4,~8,~12~$Myr after
               SN explosion with energy $E_{SN} = 10^{53}$~erg 
               in halo with mass $M = 10^7~\msun$ 
               for the spin parameter
               $\beta = 0$ -- {\it upper} and
               $\beta = 0.17$ -- {\it low} row of panels.
                }
  \label{Fig2}
\end{figure*}
In this section we perform a comparative study of the long-term 
evolution of supernova remnants in the non-rotating and 
mildly rotating galaxies. 
%Although the initial configuration of gas
%in model~1 is spherically symmetric (due to the assumed spherical symmetry
%of the dark matter halo), 
We note that the equilibrium configuration of gas is not preserved
exactly during our numerical simulations. A gradual drift from the initial equilibrium
(though very small) introduces a seed perturbation to the gas density, which is necessary 
for the instabilities to grow. These perturbations can be regarded as initial
gas inhomogeneities that are always present in protogalaxies before supernovae explosions.

Figure~2 presents snapshots of the gas density distribution in 
model~1 (top row) and model~2 (bottom row) at four consecutive
times after the supernova explosion. 
In the non-rotating model~1, a thin shell
of compressed gas separating the hot supernova ejecta from the unperturbed
gas is clearly seen at 
$t=1$~Myr. The shock wave has decelerated by
this time and its position coincides with the position of the shell.
On the contrary, in the rotating model~2 the shock wave lies ahead of the
shell at the same evolutionary time. This is a consequence of 
a lower gas density  in the center of model~2 (see dashed and 
dot-dashed lines in Fig.~\ref{Fig1}). It takes a longer time for the 
shock wave to decelerate and merge with the shell in model~2. 
The shock front in model~2 is profoundly elongated
along the vertical axis due to a non-spherical initial distribution 
of gas. 

When the characteristic cooling time becomes shorter than the 
dynamical time (the age of a supernova remnant), 
an expanding shell becomes unstable  to the Rayleigh-Taylor and 
Kelvin-Helmholtz instabilities \citep{gull}.
As a result, small ripples that distort a spherical shape of 
the shell appear in model~1 at $t=1$~Myr.
Moreover, because the shell is radiative,
we expect that the thermal instability plays a non-negligible role
in the formation of the ripples.
The subsequent evolution of the shell is governed by the 
Rayleigh-Taylor instability, which acts mostly in 
the compressed gas of the shell outside the interface 
between hot supernova ejecta and the shell of compressed material. 
The ejecta itself experiences only large 
scale distortions, which seem to be inefficient in mixing the metals 
throughout the shell. 
The characteristic time for the development of the Rayleigh-Taylor
instability is shorter for steeper initial gas density profiles 
and vice versa.
Figure~\ref{Fig1} indicates that both the gas density distribution in model~1
and the {\it vertical} gas density distribution in model~2 have 
profiles similar to $r_\ast^{-2}$, where  $r_\ast=(r^2+z^2)^{1/2}$
is the distance from the galactic center. On the other hand, 
the {\it radial} gas density distribution in model~2 is noticeably 
shallower and follows an $r_\ast^{-1.7}$ 
profile. Hence, we expect the Rayleigh-Taylor instability to grow faster
in the non-rotating model~1. This is indeed seen in the top row of 
Fig.~\ref{Fig2} -- the shell has lost its 
spherical shape by $t=4$~Myr and prominent spurs (or fingers) start 
to grow into the unperturbed medium. Model~2 shows little spurs 
at the same evolutionary 
time, though the shell has already started to show first signs of instability.
In the end of numerical simulations, model~1 develops considerably 
longer spurs than model~2. However, the number of the spurs is 
smaller in model~1 due to a larger characteristic length of 
instability. We note that the Rayleigh-Taylor instability is expected
to grow faster in protogalaxies with cuspy dark matter halos due to
a steeper gas density profile than in protogalaxies with halos 
characterized by a flat central region. However, this effect may be present
only in massive protogalaxies \citep{kitayama05}, because steep gas 
density profiles in low-mass galaxies with cuspy dark matter halos 
are likely to be destroyed by ionizing radiation from the first stars.

%We would like to mention here about Rayleigh-Taylor 
%instability growth in protogalaxies with Navarro-Frenk-White profile. Obviously,
%the instability should grow faster than in the considered models, but
%the steep profile of gas in low-mass protogalaxies with cuspy profile 
%are destructed by ionizing radiation.
%However, massive protogalaxies might be interested to study the instability
%development, the gas confiment as concluded by \citet{kitayama05} may be %not so strong.}

Figure~\ref{Fig2} shows that at $t = 12$~Myr some spurs are found 
outside the virial radius (520~pc) in the both models.
This implies that a fraction of the baryonic mass is blown away
by the supernova explosion. To calculate this fraction ($f_{\rm out}$),
we notice that the spurs are characterized by a systematically lower 
temperature than the unperturbed medium. We use this property and 
find $f_{\rm out}$ by summing up the gas mass in the computational cells
occupied by the spurs at $t=12$~Myr. We count only those computational 
cells that lie outside the virial radius.
As the spurs cross the virial radius, they sweep up some of 
the pristine gas. Since we are only interested in the blown-away gas, 
we subtract the input from this unperturbed gas. 
The resulted fraction $f_{\rm out}$ as a function of time
is shown in Fig.~\ref{Fig2a} by the solid (model~1) and dashed 
(model~2) lines. It is evident that the rotating protogalaxy (model~2) 
losses roughly twice as much baryonic mass as the non-rotating 
protogalaxy (model~1). This is a consequence of 
a shallower initial gas density profile in the rotating protogalaxy. 

The spurs have a complicated internal structure. The central regions (or
cores) of the spurs are characterized by temperatures lower than $1.5 \times 10^3$~K
and densities roughly ten times larger than those of the 
neighbouring unperturbed gas. Because the spurs move almost 
ballistically through the unperturbed gas, bow shocks form around them, 
creating envelopes of shocked gas. The envelopes are 
hotter than the cores and are characterized by  temperatures about 
$10^4$~K and densities roughly four times larger than those of 
the neighbouring unperturbed gas.

Is the blown-away gas lost to the parent galaxy? Figure~\ref{Fig4} 
shows the gas velocity field superimposed on the gas density 
distribution at $t=12$~Myr. We find that the spurs are characterized 
by mean mass-weighted velocities of the order of 
$26$~km~s$^{-1}$ in 
model~1 and $22$~km~s$^{-1}$ in model~2. In the latter model, higher 
velocities are found near the vertical axis but it may be a numerical 
artifact due to the axisymmetric nature of our numerical simulations. 
The spurs appear to have already escaped the galaxy. Indeed, the escape 
velocity at the virial radius in both models is $v_{\rm e}=(2 g_{\rm r} 
r_{\rm v})^{1/2}\approx 13$~km~s$^{-1}$, 
where $g_{\rm r}=G M_{\rm v}(r_{\rm v})/r_{\rm v}^2$ is the radial 
gravitational acceleration at 
the virial radius and $M_{\rm v}(r_{\rm v})$ is the halo mass inside 
$r_{\rm v}$.
It is obvious that the mean mass-weighted velocity of the 
spurs is at least twice as large as the escape velocity at the virial 
radius, which implies that the spurs might have escaped our protogalaxy. 

\begin{figure}
  \resizebox{\hsize}{!}{\includegraphics{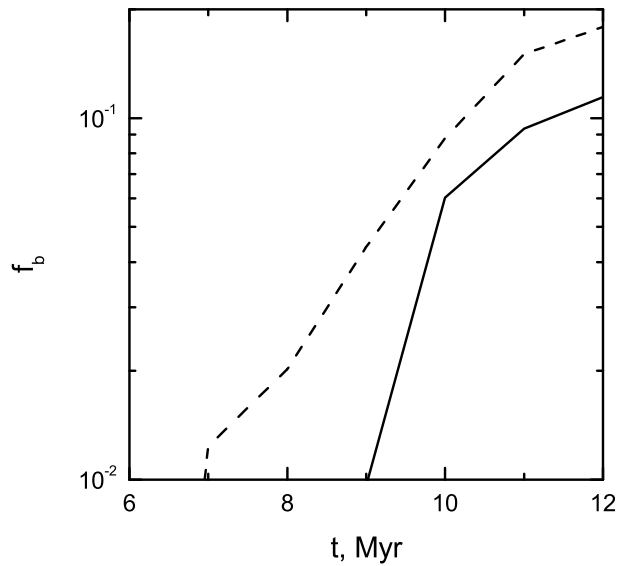}}
      \caption{Fraction of the baryonic mass blown 
               away by a supernova energy release of $10^{53}$~ergs
               in model~1 (solid) and in model~2 (dash).}
  \label{Fig2a}
\end{figure}

\begin{figure*}
  \resizebox{\hsize}{!}{\includegraphics{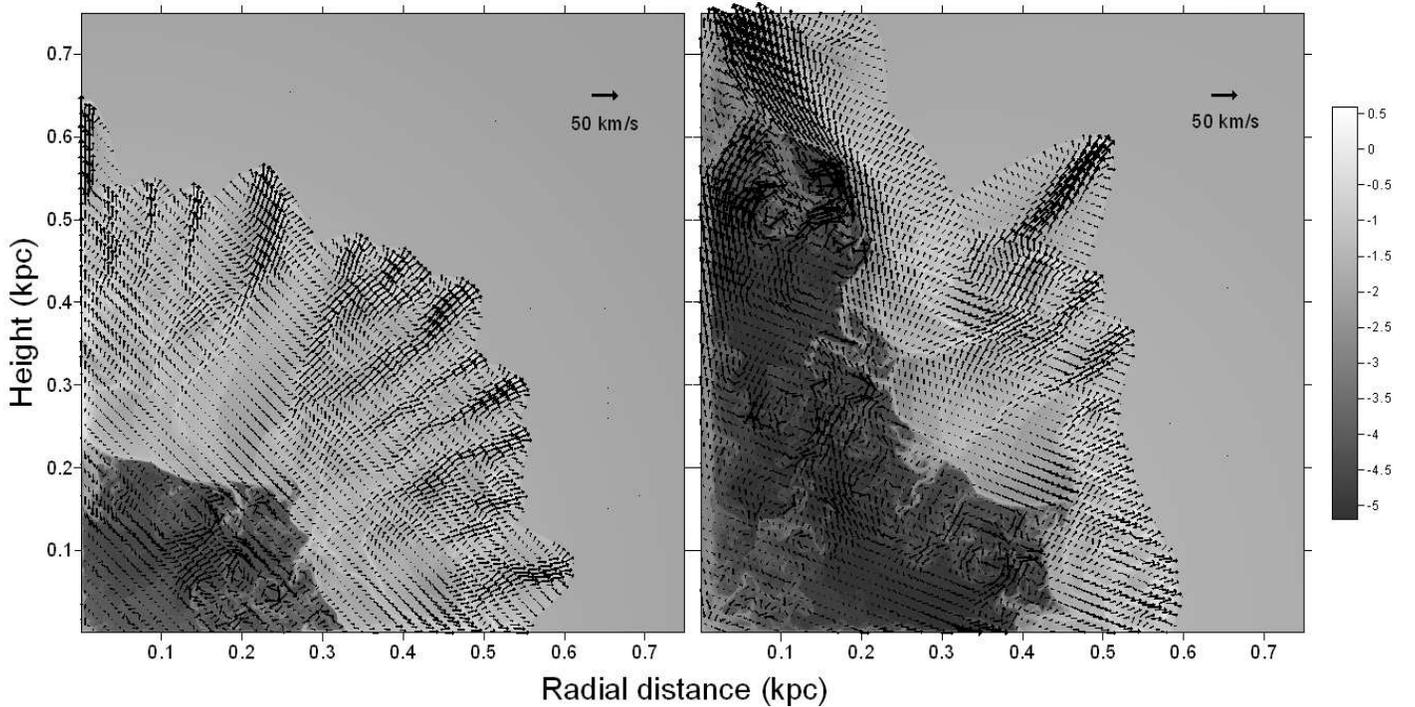}}
      \caption{
               Velocity field overlaid on the logarithmic density map.
               {\it Left panel}: protogalaxy without rotation, 
               {\it right}: with rotation. The arrow in the upper-right corner 
               corresponds to the value of velocity $50$~km~s$^{-1}$. 
                }
  \label{Fig4}
\end{figure*}

\begin{figure*}
  \resizebox{\hsize}{!}{\includegraphics{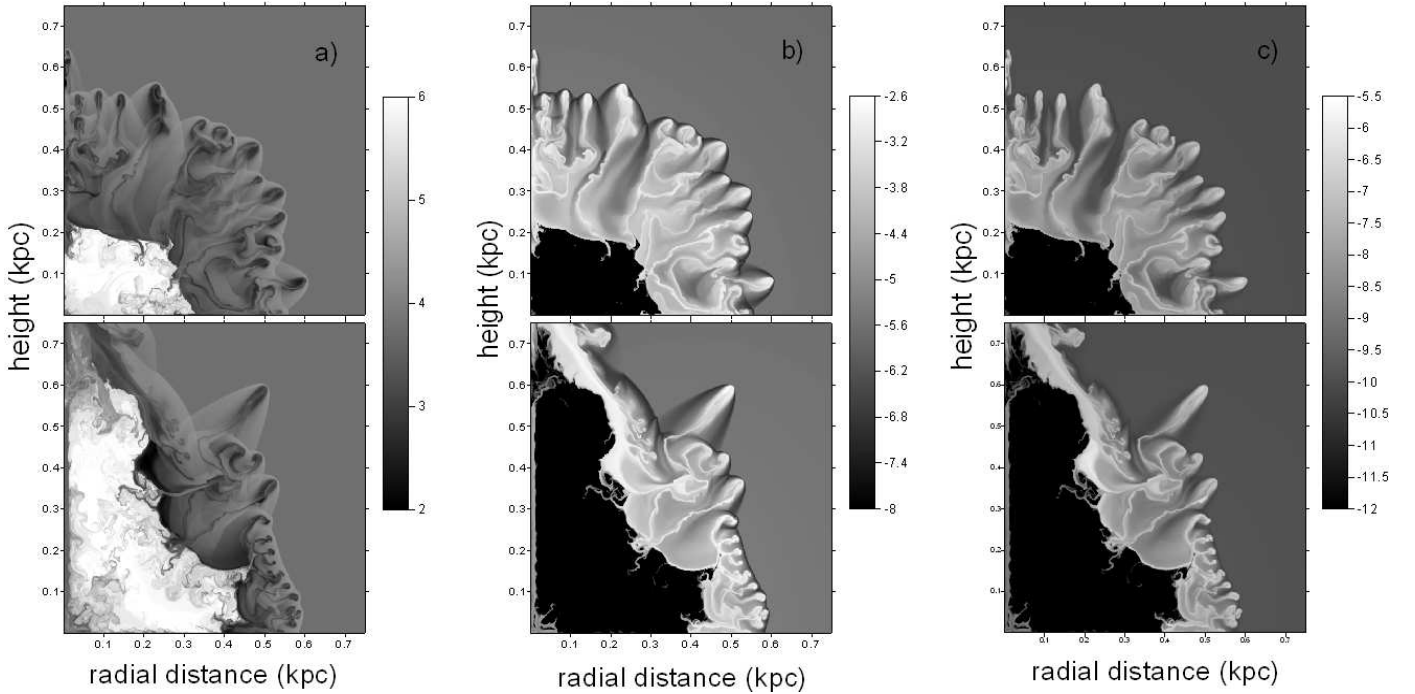}}
      \caption{
               Logarithmic distribution of 
               temperature ({\it left panel}),
               H$_2$ ({\it middle}) and
               HD ({\it right}) abundances
               at time $t=12$~Myr after the explosion of 
               SN with energy $E_{SN} = 10^{53}$~erg 
               in halo with mass $M = 10^7~\msun$ and the 
               parameter $\beta = 0$ 
               (upper panels) and $\beta = 0.4$ (lower panels). 
                }
  \label{Fig3}
\end{figure*}

%\textcolor{blue}{
%When a supernova-driven thin shell becomes radiative, its further 
%evolution is governed by cooling and destructive effect of the 
%Rayleigh-Taylor instability. In the early evolutionary phase when 
% 
%The gas in the thin shell
%cools efficiently, the temperature in some regions falls down to 100~K,
%the density becomes higher than 0.1~cm$^{-3}$. Such effective cooling
%is provided by molecular hyrogen and HD molecules, which relative
%number densities are about $10^{-3}$ and $10^{-6}$, respectively.
%That corresponds to the evolution period, when the thin shell cools
%rapidly, but it has not yet destructed enough by instabitlity,
%for instance, the snapshot for model~2 at $t=4$~Myr on Figure~\ref{Fig2}
%fits the described situation. The Jeans mass for such gas is higher 
%than the typical radial size of cold regions, which is an order of 
%thickness of the thin shell. Further development of instabilities 
%is favour to destruct gradually 
%those regions due to formation of local shock waves and, as a consequense,
%increase of temperature and destruction of HD molecules. The simple analysis
%is presented below, here we focus on the chemistry at later time, $t=12$~Myr.
%}

Figure~\ref{Fig3} shows the distribution of gas temperature (left
column), relative number densities of molecular hydrogen (middle column)
and HD molecules (right column) in the end of numerical simulations at 
$t=12$~Myr. The top/bottom rows correspond to model~1/model~2.
It is evident that low gas temperatures (below $10^3$~K)  
are found in the shell and the spurs, where cooling takes place due to 
H$_2$ and HD molecules. In particular, the lowest temperatures 
found in the spur cores are of order 500~K. 
%% Model~1 can achieve lower gas temperatures 
%% due to its higher initial gas density in the inner parts of the galaxy.
%   The middle and right columns in Fig.~\ref{Fig3} indicate that 
The relative number densities 
of molecular hydrogen and HD molecules in the spur cores 
are quite large, approximately $10^{-3}$ and $10^{-7}$, respectively. 
We note that the quoted H$_2$ relative number density is 
close to a so-called ``freeze-out'' value \citep{ohhaiman}.
A sharp decrease in the H$_2$ relative number density in the 
envelopes is explained by efficient destruction of H$_2$ molecules due 
to collisions with hydrogen atoms via the following reaction: ${\rm H_2 + H\to 3H}$. Shock waves produced by supersonic spurs heat the 
compressed gas to temperatures above $10^4$~K, at which the collisional 
destruction of molecular hydrogen becomes dominant.
There is no significant difference in the temperatures and relative number
densities of molecular species found between the two 
models. That is not unexpected because strong shock waves lead 
to similar final temperatures and 
relative number densities of species in the gas.

\begin{figure}
  \resizebox{\hsize}{!}{\includegraphics{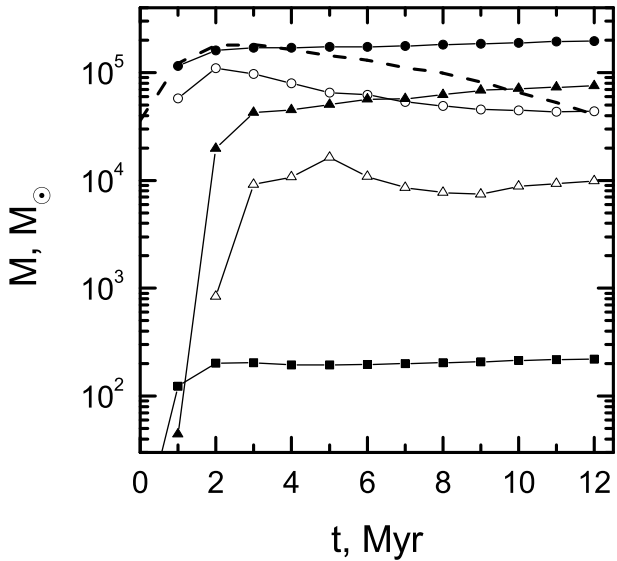}}
      \caption{
               The total mass of gas in computational domain for model~1
               with H$_2$ abundance higher than 
               $x[{\rm H_2}] = 5\times 10^{-4}$ (filled circles), 
               $x[{\rm H_2}] = 10^{-3}$ (open circles),
               with temperature lower than
               $T\leq 10^3$~K (filled triangles),
               $T\leq 500$~K  (open triagles). 
               The line with filled squares represents the total
               mass of molecular hydrogen.
               The dashed line shows the gas mass 
               contained in fragments with density
               ${\rm log}~n > -0.25$ and temperature $T<5\times10^3$ K.
                }
  \label{Fig5}
\end{figure}

\begin{figure}
  \resizebox{\hsize}{!}{\includegraphics{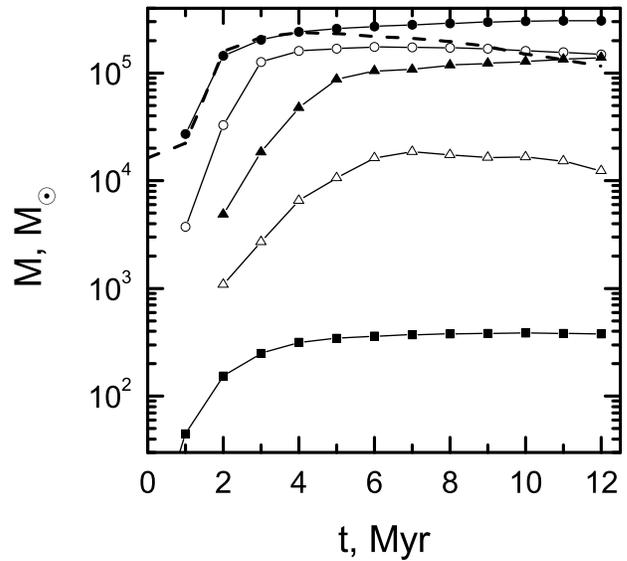}}
      \caption{
               The same as in Fig.~6 for model~2.
                }
  \label{Fig6}
\end{figure}

Figures~\ref{Fig5} and \ref{Fig6} present the temporal evolution of 
different molecular hydrogen tracers in model~1 and model~2, respectively. 
In particular, filled squares show the total molecular hydrogen mass $M_{\rm H_2}$, 
the filled and open circles yield the total mass of gas with the 
relative number density of molecular hydrogen $x[{\rm H_2}]\ge 5\times 10^{-4}$
and $x[{\rm H_2}] \ge 10^{-3}$, respectively. In addition, we keep track of the 
gas mass with temperature $\le 10^3$~K (filled triangles) and $\le 500$~K (open trinagles).
All molecular hydrogen tracers are calculated inside the computational domain.

The comparison of Figs.~\ref{Fig5} and \ref{Fig6} shows that
the H$_2$ traces saturate during the evolution.
%%, while others reach a maximum value and decline afterwards.
The saturation is explained by the fact that we consider
the gas evolution behind strong shock waves, where the H$_2$ relative density rapidly
reached a ``universal'' value $\sim 10^{-3}$ (see section 4).
The saturation times in model~2 are systematically longer than
in model~1. This can be attributed to longer cooling times in model~2
due to a shallower initial gas density profile. 
The filled circles in Figs.~\ref{Fig5} and \ref{Fig6} indicate
that the total mass of gas with $x\rm{[H_2]} \ge 5\times 10^{-4}$
saturates at $(2-3) \times 10^5~\msun$, which corresponds
to $\simeq 10\%$ of the total baryonic mass in our model galaxy.
The total gas mass with $x[{\rm H_2}] \ge 10^{-3}$ (open circles) saturates
at $\sim 10^5~\msun$. The gas mass with temperatures below 500~K
approaches an upper limit of $\sim 2\times 10^4~\msun$.
The total mass of molecular gas saturates at approximately
$300-400~\msun$, which is close the estimates made by \citet{ferrara98}.
It is also evident that the H$_2$ tracers in model~2 are
characterized by saturated values that are systematically
larger than in model~1. This is due to the fact that 
the gas mass that crosses the shock wave front is larger in model~2 than in model~1. 
%\textcolor{red}{\bf We note that a mild decline 
%(seen for some of the tracers in the late evolution) are 
%explained by outflows through the outer boundary of our 
%computational domain.}

% We note that local minima and a mild decline 
% (seen for some of the tracers in the late evolution) are 
% explained by outflows through the outer boundary of our 
% computational domain. The total gas
% loss due to these outflows amounts to $1.6\times 10^4~\msun$ 
% in model~1 and $6.3\times 10^3~\msun$ in model~2 at $t=12$~Myr.

\section{Discussion}

We have considered a SN explosion in protogalaxies with and without 
rotation, and found that rotation changes geometry and dynamics of 
the flow after the explosion, but does not affect significantly 
gas chemistry of the final state. Some differences can be found in 
temporal evolution, however the saturated values of molecular
abundances in all considered models in the two cases 
lie in a very narrow range. It is quite expected that protogalaxies with  
the ratio of rotational to gravitational energy larger than that in model~2 ($\beta=0.17$) 
are strongly violated in the vertical direction 
by radiation from the progenitor of a SN. This is basically connected with the fact that the central gas density decreases with the parameter $\beta$ 
approximately as 
$n_c\sim \exp(-\beta)$, and the characteristic size 
of a photo-ionizationally violated region varies as $R\sim n_c^{-2/3}$.  Moreover, disks in such halos are more 
extended in the radial direction with a flat radial density profile
\be
n\sim r^{-A},
\ee 
where $A\simeq -2\beta+2$, versus 
\be
n\sim z^{-2}, 
\ee 
in the vertical direction. This means that photoionizing radiation 
propagates and evacuates gas mostly perpendicular to the plane around 
the galactic center. As a result, when a SN explodes, its shell will 
expand through this cave, such that 
a significant fraction of gas can be blown away into the intergalactic medium. One can therefore speculatively conclude that the effects from 
ionizing photons of the progenitor of a SN will be more pronounced in 
rotating galaxies in the form of jet-like outflows.  

\subsection{Evolution of fragments}

One can see from Figs.~\ref{Fig2} and \ref{Fig3} that the typical 
radial length of the most cold and dense regions is several parsecs. 
The density of such clumps is about an order of 
magnitude higher than that of the surrounding gas, whereas the 
temperature is about an order lower, thus the clumps are 
in pressure equilibrium with the environment.
The clumps form due to desintegration of the shock wave under Rayleigh-Taylor 
instability which develops when gas behind the shock front starts cooling  rapidly and the front decelerates. A typical size of fragments
is expected to be close to the thickness of the compresses gas behind the front at the moment,
when it becomes unstable. 
 
Among possible mechanisms of cloud destruction 
stripping of the external layers of clouds seems the most efficient 
under the conditions of interest. This process operates mostly by 
Kelvin-Helmholtz instability \citep[e.g.,][]{klein} continuing 
on shorter scales of a single cloud. 
The typical stripping time is therefore of the order
of Kelvin-Helmholtz instability $t_{KH}^{-1} = kv/\chi^{1/2}$ \citep{chandra},
where $\chi$ is the density ratio between the cloud and 
the background medium. For typical parameters 
$v = 20$~km~s$^{-1}$, $\chi \sim 10$ and the wavelength of the order of 
the clump radius $a_0 \sim 10$~pc, the destruction time is 
$\sim 3$~Myr. This is short compared to the dynamical time, 
and from this point of view dense clumps should be destroyed quickly. 
However, the radiative cooling time is of the same order 
$t_c\sim 1-3$~Myr, which means that the density increase always 
connected with the radiative cooling can inhibite the destruction 
through stripping, so that the clumps can survive on longer dynamical 
time. The clumps however asymptotically are destroyed,  
which is seen from the fact that the mass contained in relatively dense 
($n>0.56$ cm$^{-3}$) and cold ($T<5\times 10^3$ K) fragments decreases at $t>3$ Myr as shown 
in Fig. 6 and 7 by dashed lines. Moreover, a typical mass of the most 
dense ($n>1.8$ cm$^{-3}$) clumps $M\sim 0.1-10~\msun$ (see Fig. 8) 
is much smaller than the Jeans mass for the corresponding conditions 
($n \simlt 1.8$~cm$^{-3}$ and $T \sim 0.5-1\times 10^3$~K). All this  
means that protostellar clouds do not form in the shell unless 
the clumps merge. 

\begin{figure}
  \resizebox{\hsize}{!}{\includegraphics{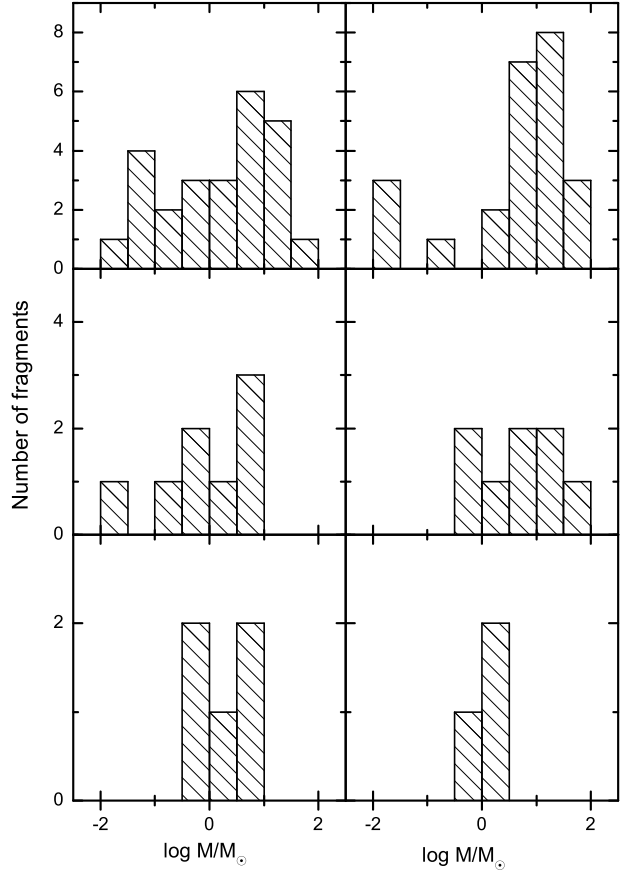}}
      \caption{
               Number of fragments in the expanding shell at 
               $t = 12$~Myr for a protogalaxy without (left) and with 
               (right) rotation; panels from the uppermost to the lowermost 
               correspond to the fragments with  ${\rm log}~n>-0.25,~0,~0.25$, 
               respectively; the mass of the fragments was calculated as 
               $\bar\rho S^{3/2}$, where $S$ is the area 
               of a filamentary fragment on the ``radius-height'' plane, 
               $\bar\rho$ is the mean density in it; note, that the fragments of 
               low density do look more filamentary and irregularly shaped than 
               the dense ones, what therefore explains a higher spread of 
               masses of fragments with lower density limit. 
               }
  \label{Fig2a}
\end{figure}

In this connection it is worth mentioning the conclusion 
by \citet{salva04} that the shell formed by a SN in the conditions similar 
to that considered here can fragment only through gravitational 
instability on to larger masses. One can think that 
the two instabilities -- Kelvin-Helmholtz and gravitational -- 
complement each other, in the sense that while Kelvin-Helmholtz 
instability of an expanding SN shell breakes it on to separate 
fragments (clouds), the gravitational instability stimulates merging 
of these low-mass fragments, and essentially can result in formation of 
protostellar clouds of sufficiently large masses. This possibility  
depends, however, on interelation between the characteristic times 
of gravitational instability and destruction of the fragments 
through stripping. As seen from Fig. 4 the relative velocities between 
the fragments and the ambient gas remain asymptotically quite 
high, 20-30 km s$^{-1}$, which results in a relatively short 
stripping destruction time: $\sim 1$ Myr -- much shorter than Jeans 
time. From this point of view the process of star formation in an 
expanding SN shell meets difficulties: from one side, low-mass fragments 
originated through disruption of the shell are sub-Jeans and cannot 
form protostellar condensation, on the other side, destruction of these 
fragments by stripping prevents them to be gravitationally collected 
into massive overcritical clouds.

\subsection{Mixing of metals}

Supernovae are recognized as a factory of metals and dust, 
although many issues related to the efficiency of metal mixing 
and  production of dust particles inside the ejecta, destruction of 
dust in supernovae shells are still under discussion 
\citep{maiolino,sugerman,meikle,venka06}. 
In particular, two conflicting conclusions about the efficiency of 
dust production in SN 2003gd event by \citep{sugerman} from one 
side and by \citep{meikle} from the other have to be mentioned. 
As far as mixing of metals in a SN driven shell is concerned, 
we believe that metals mostly confined in the border between 
the hot bubble and the surrounding interstellar gas, do penetrate and 
mix in the swept-up shell very slowly.
Mixing of metals in these conditions is determined by  
Rayleigh-Taylor instability in the interface layer \citep{wang01} 
which develops on times longer than the age of the remnant. Indeed, 
for decelerating shells the criterion for Rayleigh-Taylor instability 
reads $k<1/6\Delta R$, where $\Delta R$ is the thickness of the shell, 
$k$ is wavenumber \citep{vish1,vish2}. On the other hand, the 
instability increment is $\gamma_{\rm RT}\sim \sqrt{k|a|}$, where 
$a$ is the deceleration. This gives for a radiative remnant of the 
age $t$ the restriction $\gamma_{\rm RT}t<\sqrt{2}$, which means 
that the perturbations amplitude can grow at most by factor of $e^{\sqrt{2}}$ 
\citep{ryab}, and therefore, mixing of metals through the whole 
volume of the swept-up gas seems to be rather inefficient. From the 
point of view of our simulations this means that the effects of metals 
on cooling and chemistry of the swept-up gas can be neglected in the 
initial expansion $t\simlt 1-3$ Myr. Later on, at $t\simgt 3$ Myr, 
the shell enters an 
accelerating phase so that Rayleigh-Taylor instability grows faster. 
However, the interface between the shell and the metal enriched ejecta 
remains weakly distorted on large spatial scales, and therefore the 
metals remain mostly locked in a relatively narrow layer around the 
interface.

\section{Conclusions}

In this paper we have considered numerically the effect of energetic supernovae  
explosions ($10^{53}~$erg) in non-rotating and rotating protogalaxies with the 
total mass $10^7~\msun$ at a redshift of $z = 12$. The assumption of axial 
symmetry allowed us to evolve the supernova driven shell for several tens of 
million years. Specifically, we find the following. 

\begin{itemize}

\item The process of the shell destruction is 
different for non-rotating and rotating protogalaxies. Fingers or spurs formed 
due to various instabilities (mainly, due to the Rayleigh-Taylor instability) 
grow faster and can be found on larger distances in non-rotating galaxies. That 
is explained by a steeper initial gas density profile in non-rotating 
protogalaxies and, as a consequence, by a faster characteristic time-scale for 
the development of the Rayleigh-Taylor instability. 

\item The supernova evacuates a significant portion of the initial 
baryonic mass to radial distances comparable with or larger than 
the virial radius. 
At $t=12$~Myr after the supernova explosion 
some of the fragments of the destructed shell have attained 
mean mass-weighted velocities of the order of $30$~km~s$^{-1}$ and 
$25$~km~s$^{-1}$ in non-rotating and  rotating protogalaxies, respectively. 
These values are at least 
twice as large as the escape velocity at the virial radius, 
which implies that 
the fragments eventually may escape the protogalaxy. We estimate the 
fraction of this blown-away gas in the end of our numerical simulations 
to be about $10\%$ and $20\%$ of the initial gas mass for models without 
and with rotation, respectively.
The re-collapse phase of the shell sets in faster in models without 
rotation than in models with rotation.
These results are mostly consistent with the intuitive expectations as far as 
the difference between models with and without rotation is concerned.

\item The relative number densities of molecular hydrogen and HD molecules in 
the fingers and spurs are found to be quite large, 
approximately $10^{-3}$ and $10^{-7}$, respectively. 
The typical temperature in the spur cores is of the order of 
500~K at $t\simgt 8$~Myr after the supernova explosion. 

\item The total gas mass with $x[{\rm H_2}] \ge 10^{-3}$ 
saturates at $2 \times 10^5~\msun$, which corresponds to 
approximately $10\%$ of the total baryonic mass in our model 
galaxy. The corresponding total mass of molecular hydrogen is about $300-
400~\msun$. The saturation time-scales in the model with rotation are 
systematically longer than in the model without rotation. This can be 
attributed to longer characteristic cooling times in the model with 
rotation due to its shallower initial gas density profile.

\end{itemize}

Finally, we would like to note that the typical masses of most fragments 
(assuming their spherical symmetry) are $\sim 0.1-10~\msun$. For the typical 
densities and temperatures in the fragments to be $\sim 0.5-1$~cm$^{-3}$ and
$(0.5-1)\times 10^3$~K, respectively, 
these masses are strongly sub-Jeans and the fragments are expected to be pressure-supported. 
Their further evolution depends on both the efficiency of cooling and destruction 
due to the Kelvin-Helmholtz instability. We do not expect that the low-mass stars
can be formed in such conditions as suggested by MacKey et al. 2003, and 
Salvaterra et al. 2004. In our opinion, a more feasible mechanism 
for low-mass, metal-poor star formation is related with the re-collapse 
of a supernova bubble in protogalaxies, whose total gravitational binding energy 
is much larger than supernova energy.

\section{Acknowledgments}
We ack\-now\-le\-dge critical remarks by the anonymous referee.
We acknowledge Eugene Matvienko for his program of statistical processing. 
This work is supported by the RFBR (project codes 06-02-16819 and 08-02-91321).
E.O.V. and Yu.A.S. acknowledge partial support from
the Federal Agency of Education (project code RNP 2.1.1.3483)
and Rosnauka Agency grant No 02.438.11.7001. The simulations were done
on the Shared Hierarchical Academic Research Computing Network (SHARCNET)
while E.I.V. was a CITA National Fellow at the University of Western Ontario.

\end{document}